\documentclass[iop]{emulateapj}
\usepackage{epsfig}
\usepackage{apjfonts}
\usepackage{amsmath}
\usepackage{xcolor}
\usepackage[normalem]{ulem}
\usepackage[breaklinks,colorlinks,urlcolor=blue,citecolor=blue,linkcolor=blue]{hyperref}

\newcommand{\BNS}{S190425z\,}

\begin{document}

\title{On the energetics of a possible relativistic jet associated with the binary neutron
star merger candidate \BNS}

\author{M. Saleem $^{1}$ , L. Resmi$^{2}$, K. G. Arun$^{1}$, S. Mohan$^{2}$}
\affiliation{
$^1$ Chennai Mathematical Institute, Siruseri, India\\
$^2$Indian Institute for Space Technology, Thiruvananthapuram,
India\\
    }

\begin{abstract}
Advanced LIGO and Virgo detectors reported the first binary neutron star merger
candidate in the third observing run, \BNS,  on 25th April 2019. A weak
$\gamma$-ray excess was reported nearly coincidentally by the INTEGRAL
satellite which accidentally covered the entire localization region of
AdvLIGO/VIRGO. Electromagnetic follow-up in longer wavelengths
has not lead to the detection of any associated 
counterparts. Here we combine the available
information from gravitational wave measurements and upper limits of
fluence from \textit{INTEGRAL} to show that the observations are
completely consistent with a relativistic Gaussian structured jet and
a typical short duration Gamma Ray Burst (GRB) being produced in the merger.
We obtain posterior bounds on the on-axis isotropic equivalent energy of the associated GRB under different prior distributions. This study demonstrates  that even limited GW and EM information could be combined to produce valuable  insights about outflows from mergers. Future follow ups
may help constrain the jet structure further, especially if there is an
orphan afterglow detection associated with the candidate.
\end{abstract}

\section{Introduction}
The joint detection of GW170817~\citep{GW170817} and
GRB170817A~\citep{GRB+GW-2017} established the long standing hypothesis
that short GRBs are powered by binary neutron star (BNS) mergers~\citep{Narayan92}.
However, GRB170817A was
several orders of magnitude fainter than its cosmological counterparts \citep{Goldstein2017b}.
This led to a proposition that jets need not  successfully emerge from
some, if not all, BNS mergers and the low-energy $\gamma$-ray emission
and the non-thermal afterglow could be the result of a sub-relativistic
cocoon originating from the tidally ejected merger debris \citep{Kasliwal:2017ngb, Hallinan2017b, Gottlieb:2017pju}. 
However, late VLBI observations of GRB170817A provided a strong evidence
for the relativistic nature of the outflow \citep{Mooley:2018dlz, 2019Sci...363..968G}.  In addition, temporal
evolution of broad-band afterglow emission showed excellent agreement
with emission from a relativistic jet with an angular structure in
energy and velocity \citep{Margutti2018a, Lazzati2017a, Lyman2018a, DAvanzo:2018zyz, Resmi:2018wuc, Lamb:2018qfn}.  The low inferred energy of GRB170817A could be successfully explained by structured relativistic jet models \citep{Kathirgamaraju:2017igg,Resmi:2018wuc}. Numerical simulations of the relativistic jet piercing through the merger ejecta have shown that it successfully emerges with an angular structure \citep{2018ApJ...863...58X,2019MNRAS.484L..98K, Geng:2019qvn}. Yet, the possibility of the $\gamma$-ray emission from GRB170817A being intrinsically faint and resulting from a cocoon shock break out can still be debated \citep{2018MNRAS.475.2971B, 2018MNRAS.477.2128H}.  

Future multi-messenger observations of BNS mergers would help us answer
several open questions related to the phenomenon which include: Do all the BNS mergers
produce relativistic jets and short GRBs similar to the cosmological
sample?  If not what are the factors that determine the relative
fraction between the population which successfully launches a jet and
the one which does not?
Ongoing and future observing runs of advanced LIGO and Virgo
interferometers hence play a central role in deeply understanding the
phenomenon of BNS mergers and short GRBs.

The third observing run of LIGO and Virgo gravitational wave
interferometers reported the first binary neutron star merger candidate 
\BNS on 25th April 2019~\citep{GCN1} by the real-time processing of
the data using the {GstLAL}~\citep{gstlalpaper} and PyCBC Live~\citep{pycbc-live-Nits} analysis pipelines. 
This candidate, which was coincident in the LIGO Livingston and Virgo interferometers,  has a false alarm Rate of $4.5\times10^{-13}$ Hz (about one in $10^5$
 years) from the online analysis and a probability 
of BNS to be $\geq 99\%$. The preliminary estimate of the luminosity
distance of to the source is $156\pm 41$~Mpc~\cite{GCN2}. 
The $90$\% sky localization corresponds to $7641$~Sq degrees. 

Unlike GW170817, the poor sky localization
hampered extensive electromagnetic follow up efforts of \BNS. However,
the INTErnational Gamma-Ray Astrophysics Laboratory (\textit{INTEGRAL})
serendipitously observed the entire localization region of the
AdvLIGO/VIRGO simultaneous to \BNS, and found a low signal-to-noise
short duration ($\sim 1$~s) excess $6$~s after the merger \citep{2019GCN.24169....1M}.  
Since \textit{INTEGRAL} can not provide a localization of this excess, and
since no other confident EM counterpart is discovered till date, the
spacial coincidence of the BNS merger and the \textit{INTEGRAL} source can not be
firmly established. Nevertheless, as the entire localization region of
AdvLIGO/VIRGO is covered by the satellite, these observations to the
least provide an upper limit to the fluence of any $\gamma$-ray signal
associated with the merger. The GBM on board FERMI provided flux upper limits for a part of the LIGO/VIRGO localiztaion region \citep{2019GCN.24185....1F}. \cite{Song:2019ddw} obtained the
constraints on the viewing angle of the jet from FERMI observations to be
between  $> 0.11-0.41$ radians, assuming the GW170817 jet to be quasi-universal. 
In this letter, we ask whether the \textit{INTEGRAL} observations of
\BNS  are consistent with a relativistic jet associated with this
BNS candidate. 
We combine
two observational inputs: the luminosity distance from gravitational
waves and the INTEGRAL observations (considered both as upper limit and detection), along with
 a Gaussian structured jet model parametrized by the energy, core angle, and bulk velocity. As there are no
constraints on the inclination angle $\iota$ (same as observer's viewing
angle ${\theta_v}$ when the binary orbit is not precessing due to spins) of the binary from the
gravitational wave observations yet, we use  a simulated population
of BNS mergers and use the luminosity distance estimate from GWs
together with some conservative signal to noise ratio limits to obtain a 2
dimensional constraint in the $\iota-D_L$ plane~\citep{saleem:2019} (see
also \citep{Schutz2011,Seto:2014iya} for an analytical treatment of the
problem). 

Our results show that \BNS  could have produced a successful relativistic jet and the prompt gamma ray emission could well have been missed due to relativistic de-boost. We can derive
moderate constraints, though sensitive to the prior used, on the on-axis isotropic equivalent energy of the associated GRB (or on the total energy emitted in $\gamma$-rays, while
constraints on other parameters are weak. However, the conclusion that
the presence of a structured jet is completely consistent with the observations
itself is interesting and will help us in future to study the statistical properties of BNS mergers with poor source localization.

The remainder of the paper is organized as follows. Sec.~\ref{sec:GW}
details the input from gravitational wave observations which goes in as
prior information in the analysis of \BNS, reported in
Sec.~\ref{sec:SJ}, using structured jet
model. Sec.~\ref{sec:conclusion} discusses the implications of our
findings.

\section{Constraints from LIGO-Virgo observations}\label{sec:GW}

Even before the discovery of GW170817, it has been argued that
multi-messenger observations of binary neutron star mergers -
especially the measurement of luminosity distance and inclination angle can
have profound implications for the modelling of the
associated gamma ray burst
jet~\citep{arun2014synergy,saleem2017agparameterspace}. This is because the jets in the case of BNS mergers are very
likely to be launched along the orbital angular momentum axis of the
binary which hence relates the inclination angle  with the
viewing angle of the jet. The distance and
inclination angle in the gravitational waveforms are strongly correlated
as they both appear in the amplitude of the gravitational wave
signal~\citep{cutler1994}. Hence it is ideal to obtain the two
dimensional constraints on them using the available information and then
use that to model the $\gamma$-ray emission.

We use the following information about the binary neutron star candidate
\BNS that are available from the GCN~\citep{GCN1,GCN2}:
\begin{enumerate}
	\item {It has a probability > 99\% of being a BNS merger.}
	\item {It was observed by the network of LIGO Livingston (L1)
and Virgo (V1) detectors and since the signal to noise ratio (SNR) at Virgo was below the threshold, the candidate is considered as a single detector trigger.}      
	\item {The preliminary luminosity distance estimate is given by
$D_L = 155\pm41$Mpc.} 
\end{enumerate}   
Using the above inputs, we obtain constraints on the two-dimensional
$D_L-\iota$ space as follows. We simulate a population of BNS mergers
uniformly distributed in the comoving volume with $\cos \iota$ of the
binaries being distributed uniformly between -1 and 1. The NS masses are uniformly distributed between 1-2$M_{\odot}$. 
{We then compute optimal signal to noise ratio for each one of
them using the restricted post-Newtonian waveform~\citep{cutler1994}}.
As the trigger is an L1 single-detector trigger, we assume SNR<4 at V1, following the single-detector threshold considered by GstLAL pipeline and the L1-V1 network SNR > 9 which is motivated by the fact that the network SNR of all the O1/O2 events were above > 9 \citep{gwtc-1}. 
To compute the SNRs in L1 and V1,
we used the best reported O2 sensitivities \citep{gwtc-1} of L1 and V1 as
their representative (conservative) O3 sensitivities (see
\cite{saleem:2019} for more details). From this, we
extract a sub-population  of mergers for which the luminosity distance
distribution follows a Gaussian distribution consistent with \citep{GCN2}. The 2D distribution of
$D_L-\iota$ of this sub-population is shown in Fig. 1
which we use as the prior  for studying the prompt emission
from a short gamma ray burst associated with \BNS. 

\begin{figure}
\label{fig:resultsgw}
	\centering
	\includegraphics[scale=0.5]{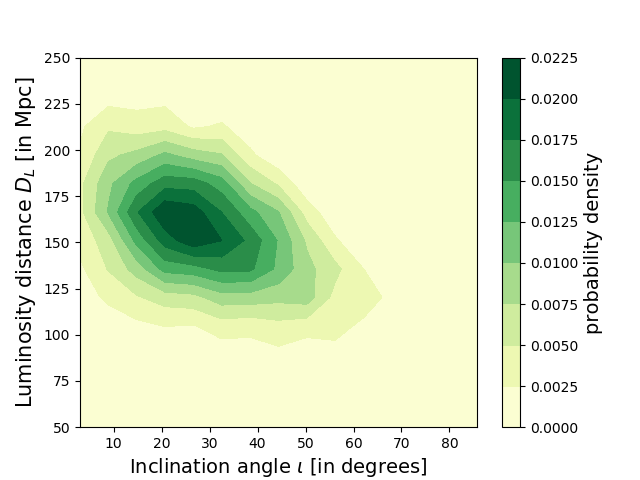}
	\caption{[Left] Constraints on the $D_L - \iota$ combination
obtained from the observed properties of \BNS as reported in
\cite{GCN2}}
\end{figure}

\section{Constraining the jet properties}
\label{sec:SJ}
In this section, we examine whether the \textit{INTEGRAL} observations are consistent with the BNS merger launching a structured relativistic jet similar to what is seen in GRB170817A \citep{LK17, Resmi:2018wuc, Lamb:2018qfn}. 

\cite{2019GCN.24169....1M} has reported a marginal ($3.7$ sigma) excess
in \textit{INTEGRAL} SPI-ACS counts temporally coincident ($+6$~s) with
the GW trigger. Such a delay can be accounted within different models of
jet ejection and $\gamma$-ray emission \citep{zhang2017}, hence it can
very likely be associated with the BNS merger candidate. However, we
treated this observation as two different possibilities. First, we
considered the fluence reported in \cite{2019GCN.24170....1M}, $(1.6 \pm
0.4) \times 10^{-7} {\rm erg}/{\rm cm}^2$ as a detection of the
associated short GRB. However, since this is a low confidence signal,
and also since its spatial coincidence to the BNS merger can not be
established, we considered $2 \times 10^{-7} erg/cm^2$ as a
conservative $3 \sigma$ upper limit to the GRB fluence. For a $\sim 1$~s
duration signal, this number is also consistent with the position
dependent sensitivity map for the duration of the GW candidate released
by the INTEGRAL collaboration \citep{2019GCN.24178....1S}, where the
fluence sensitivity ranges from $(1.5 - 6) \times 10^{-7} {\rm erg}/{\rm
cm}^2/s$. Since the \textit{FERMI} GBM has only seen about 55 percent of
the LIGO error circle \citep{2019GCN.24185....1F}, we consider \textit{INTEGRAL} observations in the rest of this paper.

Next, we computed the expected fluence from an underlying relativistic jet. The jet velocity ($\beta$) and energy have  an angular structure. The bulk Lorentz factor ($\Gamma$) distribution across the polar angle $\theta$ is given by $\Gamma \beta (\theta) = \Gamma_0 \beta_0 \exp{\frac{-\theta^2}{2 \theta_c^2}}$, where $\theta_c$ is the jet structure parameter which determines the core of the structured jet. The normalized energy profile function is given by $\epsilon(\theta) \propto \exp{\frac{-\theta^2}{\theta_c^2}}$, with the normalization constant estimated by $2 \pi \int d(\cos{\theta}) \epsilon{(\theta)} =1$. The assumed angular profile of energy and Lorentz factor are motivated by the afterglow of GRB170817A, where modelling studies have inferred such an angular profile for the outflow kinetic energy and bulk Lorentz factor \citep{Lazzati:2017zsj, Resmi:2018wuc, Granot:2017gwa, Lamb:2018qfn}. However, before extending the inferred angular profile of kinetic energy to the energy emitted in $\gamma$-rays, it must be noted that the $\gamma$-ray efficiency could have its own dependence on latitude, or the $\gamma$-ray emission mechanism could be suppressed for jet elements having low bulk Lorentz factors. We discuss these issues later in the draft.

Following the framework developed independently by \cite{donaghy05} and \cite{salafia2015structure}, the isotropic equivalent energy measured by an observer at a viewing angle $\theta_v$ is,  

\begin{equation} 
E_{\rm iso} (\theta_v)= \frac{E_{\rm tot, \gamma}}{2 \pi} \int_0^{2 \pi} d\phi \int_0^{\theta_{\rm max}} d\theta \sin(\theta) \frac{\epsilon(\theta)}{{\Gamma\!(\theta)}^4 \: \left[ 1-\beta\!(    \theta) \cos{\alpha_v} \right]^3},
\label{eq1}
\end{equation} 
where $E_{\rm tot, \gamma}$ is the total energy emitted in $\gamma$-rays, $\alpha_v$ is the angle between the the line of sight and the direction to a jet element at ($\theta, \phi$), given by $\cos(\alpha_v) =\cos(\theta_v) \cos(\theta) + \sin(\theta_v) \sin(\theta) \cos(\phi)$, and  $\theta_{\rm max}$ is the upper cut-off of integration over the polar angle of the jet. Such an upper cut-off could arise in two ways, either as the edge of the jet or as a limiting angle where the Lorentz factor or $\gamma$-ray emission efficiency drops below a certain threshold value. We numerically integrate equation-\ref{eq1} to estimate the fluence measured from the structured jet by an off-axis observer as $E_{\rm iso} (\theta_v)/4 \pi d_L^2$. 
 
Essentially, here the energy per solid angle is integrated over the jet surface after accommodating relativistic effects due to viewing angle. Therefore, this method can not reproduce time or frequency resolved quantities, such as the temporal or spectral peak in a GRB.
\begin{figure}
  \label{fig:fig1}
	\centering
	\includegraphics[scale=0.45]{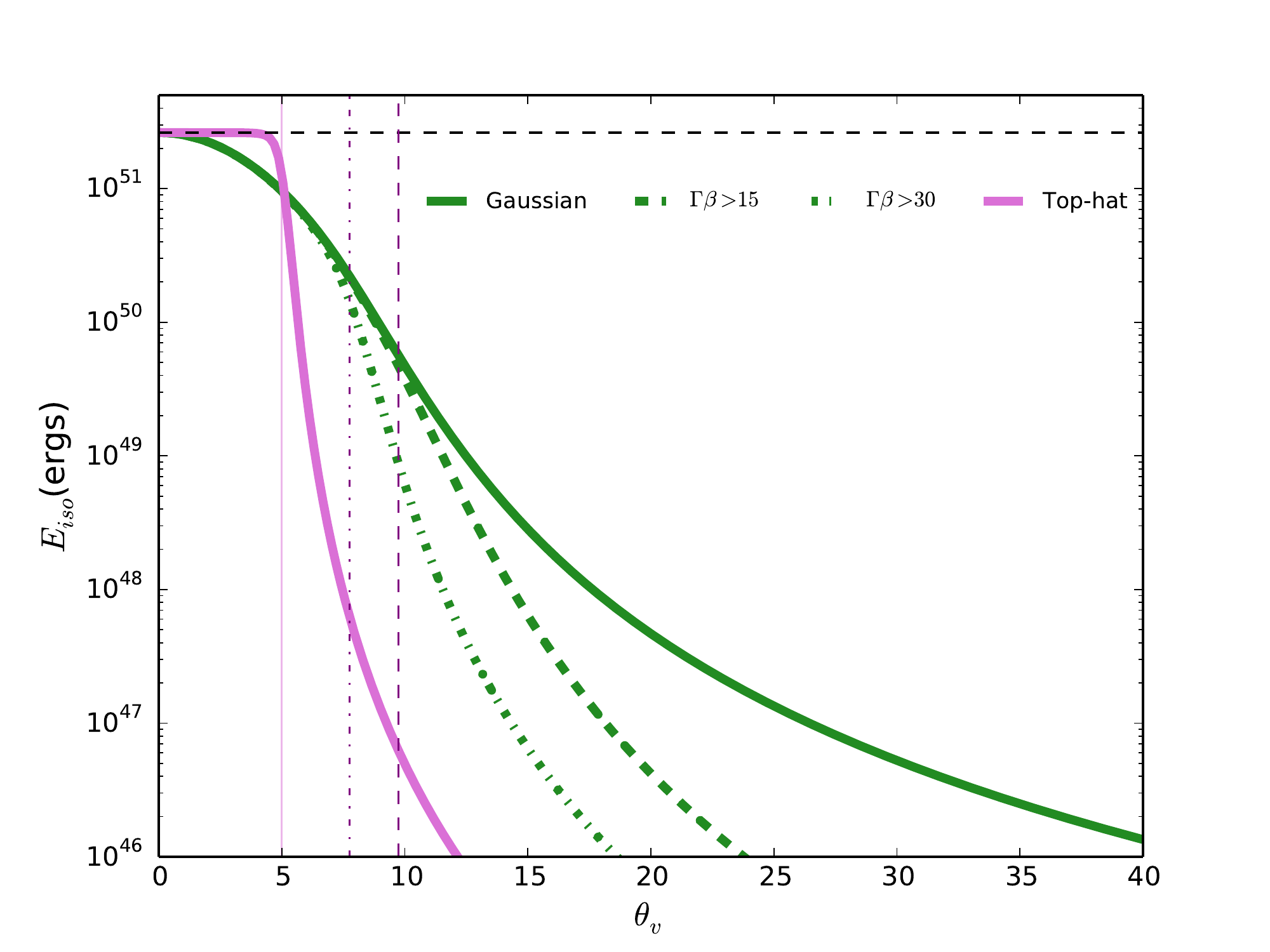}
	\caption{Variation of the isotropic equivalent energy for observers at different viewing angles, for both top-hat (violet curve) and Gaussian (green curves) jets. For both jets, we assumed $E_{\rm tot, \gamma} = 10^{49}$~ergs. Both the half-opening angle of the top-hat jet and the core-angle of the structured jet are $5^{\circ}$. Bulk Lorentz factor at the axis of the Gaussian jet is same as the bulk Lorentz factor of the top-hat jet, $100$. Horizontal dashed black line shows $E_{\rm tot,\gamma}/(1-\cos(5^{\circ})$, the on-axis $E_{\rm iso}$.  For the green solid curve the entire jet is assumed to emit $\gamma$-rays, while for the dashed and the dash-dotted green curves, the emission is restricted to a $\Gamma$ of $15$ (leading to a limit of $\theta=9.74^{\circ}$, vertical dashed line) in the integration and $30$ (leading to a limit of $7.76^{\circ}$, vertical dash-dot line) respectively.}
\end{figure}

In Fig 2, we show the behaviour of $E_{\rm iso}$ as a function of $\theta_v$ for both top-hat and Gaussian structured jets for $E_{\rm tot, \gamma} =10^{49}$~ergs. Bulk Lorentz factor of the top-hat jet, and that at the axis ($\Gamma_0$) for the Gaussian jet, are $100$. We assumed a core angle of $5^{\circ}$ for the Gaussian jet, and the same value is assumed to be he jet half-opening angle $\theta_j$ for the top-hat jet. We can see from the figure that on-axis isotropic equivalent energy for a Gaussian structured jet has the same form as that of the top-hat jet with $\theta_c$ replaced by $\theta_j$ (see \cite{sreelakshmiEtAl} for an analytical derivation of this), i.e, $E_{\rm iso}(\theta_v=0) = E_{\rm tot,\gamma}/\left[1-\cos(\theta_c)\right]$. Therefore, equation-\ref{eq1} can be rewritten in terms of $E_{\rm iso}(0)$, the isotropic equivalent energy an on-axis observer would measure. And in section-4, we present the results in terms of $E_{\rm iso}(0)$.

In Fig 2, we also show how the isotropic equivalent energy (or fluence) changes for off-axis observers if a cut-off $\Gamma$ is assumed for efficient $\gamma$-ray production in the jet. To begin with, there are not many strong observational or theoretical evidences for such a cut-off in the Lorentz factor below which $\gamma$-ray production mechanism stops.  In one example, for the low energetic GRB980425, \cite{Lithwick:2000kh} has found from optical depth arguments that a bulk Lorentz factor as low as $6.4$ is also consistent with the data. On the other hand, there are claims for a possible cut-off at relatively large Lorentz factors ($\Gamma \sim 50$) in long GRBs, estimated through statistical properties of prompt and afterglow emission \citep{Beniamini:2018udm}. It is not clear if this is applicable to short GRBs, where a structure in energy and bulk velocity can be developed as the jet propagates through the merger ejecta \citep{2018ApJ...863...58X, Geng:2019qvn, 2019MNRAS.484L..98K} Therefore, we arbitrarily assumed various cut-off Lorentz factors to see how that affects an off-axis observer. The solid green curve assumes emission from the entire jet ($\theta_{\rm max} = \pi/2$), while the dashed green curve assumes a cut-off Lorentz factor $\Gamma_{\rm cut} = 15$ and the dash-dotted green curve assumes $\Gamma_{\rm cut} = 30$. We can see that such a cut-off affects the detection at extreme viewing angles. According to the angular profile of Lorentz factor we use, if $\Gamma_{\rm cut} = \Gamma(0)/\sqrt{e}$, the emission will be restricted upto the core angle, and the Gaussian jet will behave more-or-less the same way as a top-hat jet except for a gradual decrease in fluence for $\theta_v < \theta_c$ instead of the flat profile of the top-hat jet. 

In order to better understand the constraints on a possible structured Gaussian jet, we ran a Monte-Carlo simulation with $10^5$ realizations of the jet and compared the model fluence with what is observed by \textit{INTEGRAL}. First, we used a uniform distribution in log space, ranging from $44 < \log_{10}(E_{\rm tot, \gamma}/{\rm erg})) < 51$ for $E_{\rm tot, \gamma}$. A uniform prior of $3^{\circ}< \theta_c < 20^{\circ}$ is considered for the jet core angle. With these values, we were able to cover the entire range of $E_{\rm iso} (0)$ values observed for typical cosmological short GRBs \citep{2009ApJ...703.1696Z, DAvanzo:2014urr}, and also extend the prior to much lower values if an intrinsically low energy burst is to arise from the merger. We used a wide uniform prior for the bulk Lorentz factor at the jet axis $5 < \Gamma_0 < 500$. This was done particularly because constraints on the initial bulk Lorentz factor from GRB170817A is very weak \citep{Troja2018c, Resmi:2018wuc}, and we do not have a good prior information on the kind of outflows arising from BNS mergers.  

In the next step, we chose priors that best reproduces the observed short GRB fluence distribution from Fermi 4-yr catalogue \citep{Gruber:2014iza}. We found that a broken power-law prior distribution of $E_{\rm tot, \gamma}$ along with uniform distributions $3^{\circ} < \theta_c < 20^{\circ}$ and $100 < \Gamma < 500$  are able to reproduce the observed fluence distribution above $2\times 10^{-7}$~erg/cm$^2$ relatively well   \citep{sreelakshmiEtAl}. We assumed $E_{\rm tot, \gamma}$ to extend from $5 \times 10^{47}$~ergs to $10^{50}$~ergs with a power-law index of $-0.53$ and from $10^{50}$~ergs to $5 \times 10^{51}$~ergs with an index of $-3.5$. For the indices, we adopted values from the luminosity function used by \cite{ghirlanda2016short}.

We used both these distributions along with the \textit{INTEGRAL} observations to see constraints on a possible GRB associated with the merger. The $D_L -\iota$ distribution computed in the previous section is substituted as the prior for $D_L$ and $\theta_v$.  We extracted marginalized posterior distributions for $\theta_c, \Gamma_0, \theta_v,$ and  $E_{\rm tot, gamma}$ which later we converted to the posterior of $E_{\rm iso}(0)$.

We find that the \textit{INTEGRAL} fluence provides a good constraint to the energy of the Gamma Ray Burst. The isotropic equivalent energies of cosmological short GRBs detected by FERMI GBM and SWIFT BAT ranges from $10^{48}$ to $10^{53}$~ergs \citep{2009ApJ...703.1696Z}. Our analysis shows that, had the observer been along the axis of the jet, a typical short GRB could have been detected along with \BNS(see Fig 3). The uniform energy prior, which has a wide range, get well constrained by the \textit{INTEGRAL} observations. When considered as a detection, $E_{\rm iso}(0)$ is tightly constrained to be between $(4.74 \times 10^{47} - 2.21 \times 10^{51})$~ergs (blue curve in the left panel of Fig 3). On the other hand, when considered as an upper limit, the posterior (orange curve) indicates that $E_{\rm iso}(0) \leq 3 \times 10^{48}$~ergs at $1 \sigma$ level  broadly in agreement with the range observed for standard cosmological short GRBs. The lower end of the posterior in this case is not constrained (as expected in the absence of a detection) and hence simply follows the prior on  $E_{\rm iso}(0)$ (see Fig 3, shaded grey). 

On the other hand, for the second case, where we use a narrower prior distribution in energy, the observations are not able to place tight constraints on the assumed prior distribution. The $1-\sigma$ posterior bounds is $1.9 \times 10^{49}$~erg $< E_{\rm iso}(0) < 6.6 \times 10^{50}$~erg. Though the posterior bounds are sensitive to the assumed prior, both prior distributions we considered here imply that the observations can not rule out an event with typical short GRB energetics. We recall that our conclusion is sensitive to deeper limits from \textit{INTEGRAL} as well as the refined GW posteriors on $D_L - \iota$. 

The $\gamma$-ray observations can not provide any useful constraints to the jet core angle or its initial bulk Lorentz factor. Taken either as upper limit or as a detection, the \textit{INTEGRAL} fluence is consistent with the expected emission from a relativistic jet. We also ran the simulations where $E_{\rm iso}(\theta_v)$ is calculated by assuming that the $\gamma$-ray emission stops below a bulk Lorentz factor of $15$.  The posterior distribution from such a model did not show much differences from a model where the entire jet surface is integrated to obtain $E_{\rm iso}$.

\begin{figure*}
\label{fig:results2}
	\centering
	\includegraphics[scale=0.5]{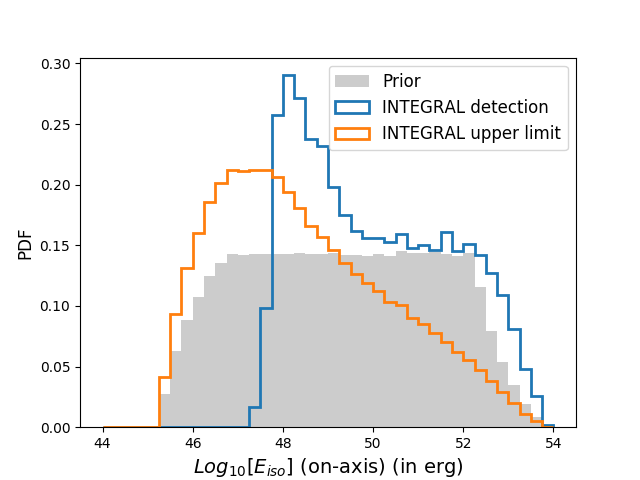}
		\includegraphics[scale=0.5]{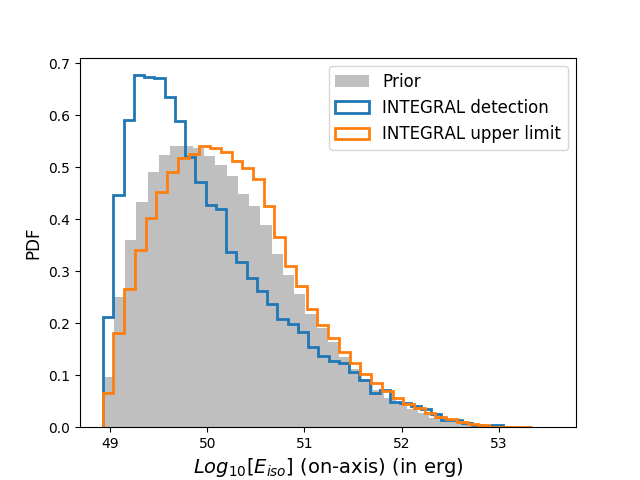}
	\caption{Constraints on the isotropic
on-axis energy of the short-GRB associated with \BNS assuming a Gaussian
structured jet. On the left panel, the grey shades indicates the prior distribution which
results from assuming uniform priors on $\log_{10} (E_{\rm tot, \gamma}/erg)$ in the range of [$44 - 51$] and on $\theta_c$ in [3,20] degrees. On the right panel, same prior is used for $\theta_c$ while for $E_{\rm tot, \gamma}$, a broken power-law function is used which reproduces the observed fluence distribution of short GRBs (see text for details). The orange curve results from considering a fluence upper limit of $2 \times 10^{-7}$ erg/cm$^2$ while the blue curve considers a  detection of $1.6 \pm 0.4$~erg/cm$^2$ as reported in \citep{2019GCN.24170....1M}. Treating the low S/N excess as a detection, the isotropic equivalent energy of an associated GRB, if viewed on-axis, is tightly constrained for a flat prior. In both cases, on-axis energy of a possible associated GRB is within the range of that of the cosmological SGRB population.}
\end{figure*}
\section{Discussion and conclusions}\label{sec:conclusion}
One of the important open questions from the observations of the first
BNS merger GW170817 was whether the gamma ray emission from GRB170817A
is powered by a relativistic jet, and whether this burst is one among
the population of cosmological short GRBs. The absence of a strong gamma
ray burst associated with \BNS, the second BNS merger candidate by
LIGO/VIRGO,  increased the importance of this question. 

While the evolution of the broad-band afterglow and proper motion of the
radio afterglow of GRB170817A confirm the presence of a relativistic
jet, the low energetic $\gamma$-ray signal could still be
 argued to be different from cosmological short GRBs. 
If so, a novel class of low luminosity gamma ray transients will be seen
associated with BNS mergers. 
Therefore, it is important to understand whether jets are associated with BNS mergers, and if so to obtain reliable constraints on their energetics. 

In this analysis, we have shown that electromagnetic observations of \BNS is consistent with the launch of a relativistic jet typical to that of short duration GRBs. We see that a structured jet with Gaussian profile at the distance of \BNS is well consistent with the INTEGRAL sensitivity limits. The inferred posterior of the isotropic equivalent energy for an on-axis observer is in agreement with that of typical short GRBs. Even when we take the conservative view that \textit{INTEGRAL} observations yielded only an upper limit, thereby allowing even low energy explosions to remain consistent, the $1 \sigma$ posterior results in $E_{\rm iso}(0) \leq 3 \times 10^{48}$~erg for a wide uniform prior in $E_{\rm tot,\gamma}$. Constraints are not very tight if a broken power-law prior distribution, which reproduces the fluence distribution of standard short GRBs, is assumed for $E_{\rm tot,\gamma}$. Nevertheless, this indicates that one need not invoke a intrinsically low energy GRB or shock breakout to explain the absence of a strong $\gamma$-ray signal above \textit{INTEGRAL} limits. We do not find any significant change in the results if $\gamma$-ray emission is assumed to happen only from those regions of the jet above a moderate threshold Lorentz factor.

A limitation of this approach to estimate off-axis fluence is that it can not accommodate energy dissipation and $\gamma$-ray production mechanisms. This approach is also not sensitive to temporal and spectral evolution of the radiation, instead it returns the total fluence observed. Our conclusions are sensitive to future deeper upper limits from \textit{INTEGRAL}. Future multi-messenger observations of Binary NS
mergers in AdvLIGO/VIRGO 3rd observing run will certainly provide further
valuable  insights into the physics of these mergers and Gamma Ray Bursts.

\section*{Acknowledgements}
K. G. A. and M. S. are partially supported by a grant
from Infosys Foundation. K. G. A. acknowledges the sup-
port by the Indo-US Science and Technology Forum through
the Indo-US Center for the Exploration of Extreme Gravity
(Grant No. IUSSTF/JC-029/2016). R.L. acknowledges support from the grant EMR/2016/007127 from Department of Science and Technology, India. We thank an anonymous referee whose suggestions greatly improved this manuscript.

\bibliographystyle{apj}
\bibliography{S190425}

\begin{thebibliography}{}
\expandafter\ifx\csname natexlab\endcsname\relax\def\natexlab#1{#1}\fi

\bibitem[{Abbott {et~al.}(2017{\natexlab{a}})Abbott, Abbott, Abbott, Acernese,
  Ackley, Adams, Adams, Addesso, Adhikari, Adya, {et~al.}}]{GRB+GW-2017}
Abbott, B., Abbott, R., Abbott, T., {et~al.} 2017{\natexlab{a}}, The
  Astrophysical Journal Letters, 848, L13

\bibitem[{Abbott {et~al.}(2017{\natexlab{b}})Abbott, Abbott, Abbott, Acernese,
  Ackley, Adams, Adams, Addesso, Adhikari, Adya, {et~al.}}]{GW170817}
Abbott, B.~P., Abbott, R., Abbott, T., {et~al.} 2017{\natexlab{b}}, Physical
  Review Letters, 119, 161101

\bibitem[{Abbott {et~al.}(2018)}]{gwtc-1}
Abbott, B.~P., {et~al.} 2018, arXiv:1811.12907

\bibitem[{Arun {et~al.}(2014)Arun, Tagoshi, Pai, \& Mishra}]{arun2014synergy}
Arun, K., Tagoshi, H., Pai, A., \& Mishra, C.~K. 2014, Physical Review D, 90,
  024060

\bibitem[{Beniamini \& Nakar(2019)}]{Beniamini:2018udm}
Beniamini, P., \& Nakar, E. 2019, Mon. Not. Roy. Astron. Soc., 482, 5430

\bibitem[{{Bromberg} {et~al.}(2018){Bromberg}, {Tchekhovskoy}, {Gottlieb},
  {Nakar}, \& {Piran}}]{2018MNRAS.475.2971B}
{Bromberg}, O., {Tchekhovskoy}, A., {Gottlieb}, O., {Nakar}, E., \& {Piran}, T.
  2018, \mnras, 475, 2971

\bibitem[{Cutler \& Flanagan(1994)}]{cutler1994}
Cutler, C., \& Flanagan, E.~E. 1994, Physical Review D, 49, 2658

\bibitem[{D'Avanzo {et~al.}(2014)}]{DAvanzo:2014urr}
D'Avanzo, P., {et~al.} 2014, Mon. Not. Roy. Astron. Soc., 442, 2342

\bibitem[{D'Avanzo {et~al.}(2018)}]{DAvanzo:2018zyz}
---. 2018, Astron. Astrophys., 613, L1

\bibitem[{{Donaghy}(2006)}]{donaghy05}
{Donaghy}, T.~Q. 2006, \apj, 645, 436

\bibitem[{Fletcher \& et~al.(2019)}]{2019GCN.24185....1F}
Fletcher, C., \& et~al. 2019, GRB Coordinates Network, Circular Service,
  No.~24185, \#1 (2019), 24185

\bibitem[{Geng {et~al.}(2019)Geng, Zhang, Kölligan, Kuiper, \&
  Huang}]{Geng:2019qvn}
Geng, J.-J., Zhang, B., Kölligan, A., Kuiper, R., \& Huang, Y.-F. 2019,
  Astrophys. J., 877, L40

\bibitem[{Ghirlanda {et~al.}(2016)Ghirlanda, Salafia, Pescalli, Ghisellini,
  Salvaterra, Chassande-Mottin, Colpi, Nappo, D’Avanzo, Melandri,
  {et~al.}}]{ghirlanda2016short}
Ghirlanda, G., Salafia, O., Pescalli, A., {et~al.} 2016, Astronomy \&
  Astrophysics, 594, A84

\bibitem[{{Ghirlanda} {et~al.}(2019){Ghirlanda}, {Salafia}, {Paragi},
  {Giroletti}, {Yang}, {Marcote}, {Blanchard}, {Agudo}, {An}, {Bernardini},
  {Beswick}, {Branchesi}, {Campana}, {Casadio}, {Chassande-Mottin}, {Colpi},
  {Covino}, {D'Avanzo}, {D'Elia}, {Frey}, {Gawronski}, {Ghisellini}, {Gurvits},
  {Jonker}, {van Langevelde}, {Melandri}, {Moldon}, {Nava}, {Perego},
  {Perez-Torres}, {Reynolds}, {Salvaterra}, {Tagliaferri}, {Venturi},
  {Vergani}, \& {Zhang}}]{2019Sci...363..968G}
{Ghirlanda}, G., {Salafia}, O.~S., {Paragi}, Z., {et~al.} 2019, Science, 363,
  968

\bibitem[{{Goldstein} {et~al.}(2017b){Goldstein}, {Veres}, {Burns}, {Briggs},
  {Hamburg}, {Kocevski}, {Wilson-Hodge}, {Preece}, {Poolakkil}, {Roberts},
  {Hui}, {Connaughton}, {Racusin}, {von Kienlin}, {Dal Canton}, {Christensen},
  {Littenberg}, {Siellez}, {Blackburn}, {Broida}, {Bissaldi}, {Cleveland},
  {Gibby}, {Giles}, {Kippen}, {McBreen}, {McEnery}, {Meegan}, {Paciesas}, \&
  {Stanbro}}]{Goldstein2017b}
{Goldstein}, A., {Veres}, P., {Burns}, E., {et~al.} 2017b, \apjl, 848, L14

\bibitem[{Gottlieb {et~al.}(2018)Gottlieb, Nakar, Piran, \&
  Hotokezaka}]{Gottlieb:2017pju}
Gottlieb, O., Nakar, E., Piran, T., \& Hotokezaka, K. 2018, Mon. Not. Roy.
  Astron. Soc., 479, 588

\bibitem[{Granot {et~al.}(2018)Granot, Gill, Guetta, \&
  De~Colle}]{Granot:2017gwa}
Granot, J., Gill, R., Guetta, D., \& De~Colle, F. 2018, Mon. Not. Roy. Astron.
  Soc., 481, 1597

\bibitem[{Gruber {et~al.}(2014)}]{Gruber:2014iza}
Gruber, D., {et~al.} 2014, Astrophys. J. Suppl., 211, 12

\bibitem[{{Hallinan} {et~al.}(2017){Hallinan}, {Corsi}, {Mooley}, {Hotokezaka},
  {Nakar}, {Kasliwal}, {Kaplan}, {Frail}, {Myers}, {Murphy}, {De}, {Dobie},
  {Allison}, {Bannister}, {Bhalerao}, {Chandra}, {Clarke}, {Giacintucci}, {Ho},
  {Horesh}, {Kassim}, {Kulkarni}, {Lenc}, {Lockman}, {Lynch}, {Nichols},
  {Nissanke}, {Palliyaguru}, {Peters}, {Piran}, {Rana}, {Sadler}, \&
  {Singer}}]{Hallinan2017b}
{Hallinan}, G., {Corsi}, A., {Mooley}, K.~P., {et~al.} 2017, Science, 358, 1579

\bibitem[{{Harrison} {et~al.}(2018){Harrison}, {Gottlieb}, \&
  {Nakar}}]{2018MNRAS.477.2128H}
{Harrison}, R., {Gottlieb}, O., \& {Nakar}, E. 2018, \mnras, 477, 2128

\bibitem[{Kasliwal {et~al.}(2017)}]{Kasliwal:2017ngb}
Kasliwal, M.~M., {et~al.} 2017, Science, 358, 1559

\bibitem[{Kathirgamaraju {et~al.}(2018)Kathirgamaraju, Barniol~Duran, \&
  Giannios}]{Kathirgamaraju:2017igg}
Kathirgamaraju, A., Barniol~Duran, R., \& Giannios, D. 2018, Mon. Not. Roy.
  Astron. Soc., 473, L121

\bibitem[{{Kathirgamaraju} {et~al.}(2019){Kathirgamaraju}, {Tchekhovskoy},
  {Giannios}, \& {Barniol Duran}}]{2019MNRAS.484L..98K}
{Kathirgamaraju}, A., {Tchekhovskoy}, A., {Giannios}, D., \& {Barniol Duran},
  R. 2019, \mnras, 484, L98

\bibitem[{Lamb \& Kobayashi(2017)}]{LK17}
Lamb, G.~P., \& Kobayashi, S. 2017, Mon. Not. Roy. Astron. Soc., 472, 4953

\bibitem[{Lamb {et~al.}(2019)}]{Lamb:2018qfn}
Lamb, G.~P., {et~al.} 2019, Astrophys. J., 870, L15

\bibitem[{{Lazzati} {et~al.}(2017){Lazzati}, {Perna}, {Morsony},
  {L{\'o}pez-C{\'a}mara}, {Cantiello}, {Ciolfi}, {giacomazzo}, \&
  {Workman}}]{Lazzati2017a}
{Lazzati}, D., {Perna}, R., {Morsony}, B.~J., {et~al.} 2017, \nat, submitted,
  arXiv:1712.03237

\bibitem[{Lazzati {et~al.}(2018)Lazzati, Perna, Morsony, López-Cámara,
  Cantiello, Ciolfi, Giacomazzo, \& Workman}]{Lazzati:2017zsj}
Lazzati, D., Perna, R., Morsony, B.~J., {et~al.} 2018, Phys. Rev. Lett., 120,
  241103

\bibitem[{LIGO \& Collaborations(2019{\natexlab{a}})}]{GCN1}
LIGO, \& Collaborations, V. 2019{\natexlab{a}}, GRB Coordinates Network,
  Circular Service, No 24168

\bibitem[{LIGO \& Collaborations(2019{\natexlab{b}})}]{GCN2}
---. 2019{\natexlab{b}}, GRB Coordinates Network, Circular Service, No 24228

\bibitem[{Lithwick \& Sari(2001)}]{Lithwick:2000kh}
Lithwick, Y., \& Sari, R. 2001, Astrophys. J., 555, 540

\bibitem[{{Lyman} {et~al.}(2018){Lyman}, {Lamb}, {Levan}, {Mandel}, {Tanvir},
  {Kobayashi}, {Gompertz}, {Hjorth}, {Fruchter}, {Kangas}, {Steeghs}, {Steele},
  {Cano}, {Copperwheat}, {Evans}, {Fynbo}, {Gall}, {Im}, {Izzo}, {Jakobsson},
  {Milvang-Jensen}, {O'Brien}, {Osborne}, {Palazzi}, {Perley}, {Pian},
  {Rosswog}, {Rowlinson}, {Schulze}, {Stanway}, {Sutton}, {Th{\"o}ne}, {de
  Ugarte Postigo}, {Watson}, {Wiersema}, \& {Wijers}}]{Lyman2018a}
{Lyman}, J.~D., {Lamb}, G.~P., {Levan}, A.~J., {et~al.} 2018, Nature Astronomy,
  arXiv:1801.02669

\bibitem[{{Margutti} {et~al.}(2018){Margutti}, {Alexander}, {Xie}, {Sironi},
  {Metzger}, {Kathirgamaraju}, {Fong}, {Blanchard}, {Berger}, {MacFadyen},
  {Giannios}, {Guidorzi}, {Hajela}, {Chornock}, {Cowperthwaite}, {Eftekhari},
  {Nicholl}, {Villar}, {Williams}, \& {Zrake}}]{Margutti2018a}
{Margutti}, R., {Alexander}, K.~D., {Xie}, X., {et~al.} 2018, \apj, submitted,
  arXiv:1801.03531

\bibitem[{Martin-Carillo \& et~al.(2019)}]{2019GCN.24169....1M}
Martin-Carillo, \& et~al. 2019, GRB Coordinates Network, Circular Service,
  No.~24169, \#1 (2019), 24169

\bibitem[{Messick {et~al.}(2017)Messick, Blackburn, Brady, Brockill, Cannon,
  Cariou, Caudill, Chamberlin, Creighton, Everett, Hanna, Keppel, Lang, Li,
  Meacher, Nielsen, Pankow, Privitera, Qi, Sachdev, Sadeghian, Singer, Thomas,
  Wade, Wade, Weinstein, \& Wiesner}]{gstlalpaper}
Messick, C., Blackburn, K., Brady, P., {et~al.} 2017, Phys. Rev. D, 95, 042001

\bibitem[{Minaev \& et~al.(2019)}]{2019GCN.24170....1M}
Minaev, P., \& et~al. 2019, GRB Coordinates Network, Circular Service,
  No.~24170, \#1 (2019), 24170

\bibitem[{Mohan {et~al.}(2019)Mohan, Saleem, \& Resmi}]{sreelakshmiEtAl}
Mohan, S., Saleem, M., \& Resmi, L. 2019, arXiv:1912.09436

\bibitem[{Mooley {et~al.}(2018)Mooley, Deller, Gottlieb, Nakar, Hallinan,
  Bourke, Frail, Horesh, Corsi, \& Hotokezaka}]{Mooley:2018dlz}
Mooley, K.~P., Deller, A.~T., Gottlieb, O., {et~al.} 2018, Nature, 561, 355

\bibitem[{{Narayan} {et~al.}(1992){Narayan}, {Paczynski}, \&
  {Piran}}]{Narayan92}
{Narayan}, R., {Paczynski}, B., \& {Piran}, T. 1992, Astrophys. J., 395, L83

\bibitem[{Nitz {et~al.}(2018)Nitz, Dal~Canton, Davis, \&
  Reyes}]{pycbc-live-Nits}
Nitz, A.~H., Dal~Canton, T., Davis, D., \& Reyes, S. 2018, Phys. Rev. D, 98,
  024050

\bibitem[{Resmi {et~al.}(2018)}]{Resmi:2018wuc}
Resmi, L., {et~al.} 2018, Astrophys. J., 867, 57

\bibitem[{Salafia {et~al.}(2015)Salafia, Ghisellini, Pescalli, Ghirlanda, \&
  Nappo}]{salafia2015structure}
Salafia, O., Ghisellini, G., Pescalli, A., Ghirlanda, G., \& Nappo, F. 2015,
  Monthly Notices of the Royal Astronomical Society, 450, 3549

\bibitem[{Saleem(2019)}]{saleem:2019}
Saleem, M. 2019, in preparation

\bibitem[{Saleem {et~al.}(2018)Saleem, Resmi, Misra, Pai, \&
  Arun}]{saleem2017agparameterspace}
Saleem, M., Resmi, L., Misra, K., Pai, A., \& Arun, K.~G. 2018, Monthly Notices
  of the Royal Astronomical Society, 474, 5340

\bibitem[{{Savchenko} \& et~al.(2019)}]{2019GCN.24178....1S}
{Savchenko}, V., \& et~al. 2019, GRB Coordinates Network, Circular Service,
  No.~23807, \#1 (2019), 24178

\bibitem[{{Schutz}(2011)}]{Schutz2011}
{Schutz}, B.~F. 2011, Classical and Quantum Gravity, 28, 125023

\bibitem[{Seto(2015)}]{Seto:2014iya}
Seto, N. 2015, Mon. Not. Roy. Astron. Soc., 446, 2887

\bibitem[{Song {et~al.}(2019)Song, Ai, Wang, Xing, Gao, \&
  Zhang}]{Song:2019ddw}
Song, H.-R., Ai, S.-K., Wang, M.-H., {et~al.} 2019, arXiv:1904.12263

\bibitem[{{Troja} {et~al.}(2018){Troja}, {Piro}, {Ryan}, {van Eerten}, {Ricci},
  {Wieringa}, {Lotti}, {Sakamoto}, \& {Cenko}}]{Troja2018c}
{Troja}, E., {Piro}, L., {Ryan}, G., {et~al.} 2018, \mnras, submitted,
  arXiv:1801.06516

\bibitem[{{Xie} {et~al.}(2018){Xie}, {Zrake}, \&
  {MacFadyen}}]{2018ApJ...863...58X}
{Xie}, X., {Zrake}, J., \& {MacFadyen}, A. 2018, \apj, 863, 58

\bibitem[{{Zhang} {et~al.}(2009){Zhang}, {Zhang}, {Virgili}, {Liang}, {Kann},
  {Wu}, {Proga}, {Lv}, {Toma}, {M{\'e}sz{\'a}ros}, {Burrows}, {Roming}, \&
  {Gehrels}}]{2009ApJ...703.1696Z}
{Zhang}, B., {Zhang}, B.-B., {Virgili}, F.~J., {et~al.} 2009, \apj, 703, 1696

\bibitem[{Zhang {et~al.}(2018)}]{zhang2017}
Zhang, B.~B., {et~al.} 2018, Nature Commun., 9, 447

\end{thebibliography}
\end{document}